\documentclass{article}

\usepackage{url} %Para las url de Bibtex.
\usepackage{graphicx}
\usepackage{natbib}
\usepackage{amsthm}
\usepackage{amsmath} %Para begin{align*}
\usepackage{amssymb}
\usepackage{float} %Para incluir las grÃ¡ficas y tablas donde queramos, opciÃ³n [H].
\usepackage{hyperref}
\usepackage{color}
\usepackage{url} %Para las url de Bibtex.
\usepackage{caption} %Personalizar las caption.
\usepackage{rotating}
\usepackage{slashbox} %Para partir una celda de una tabla en dos: comando \backslashbox
\usepackage{multicol} %Para dividir el texto en las columnas que se quiera,
\usepackage{url}
\usepackage{multirow} %Para dividir el texto en las columnas que se quiera,
\newcommand{\re}{\mathbb R}
{\bf}{\rm}
\newtheorem{thm}{Theorem}
\newtheorem{prop}[thm]{Proposition}

\newtheorem{defn}[thm]{Definition}

\newcommand{\norm}[1]{\left\lVert#1\right\rVert}

\DeclareMathOperator{\rank}{rank}
\DeclareMathOperator*{\argmin}{arg\,min}

\begin{document}

\title{On using Reproducible Hilbert Spaces for the analysis of Replicated Spatial Point Processes.}

\author{A.~Sim\'o \\
\small{Department of Mathematics-IMAC. Universitat Jaume I.
              Avda.~del Riu Sec s/n. 12071-Castell\'o, Spain.}}
 \maketitle

\begin{abstract}
 This paper focuses on the use of the theory of Reproducing Kernel Hilbert Spaces in the statistical analysis of replicated point processes. We show that spatial point processes can be observed as random variables in a Reproducing Kernel Hilbert Space and, as a result, methodological and theoretical results for statistical analysis in these spaces can be applied to them. In particular and by way of illustration, we show how we can use the proposed methodology to identify differences between several classes of replicate point patterns using the MBox and MANOVA tests, and to classify a new observation, using Discriminant Functions.

\textbf{keyword}

Reproducing Kernel Hilbert Space; Functional Data Analysis;  Replicated spatial point processes;  Analysis of Variance; Supervised Classification
\end{abstract}

%% \linenumbers

%% main text% ===============================================================================================================
\section{Introduction} \label{sec:introduccion}
% ===============================================================================================================

A spatial point process is a stochastic random process whose realizations are locally finite sets of points (locations of events) in a study region $E$ of $\re^2$. The term locally finite means  that
for any Borel bounded set there is a finite number of points (with probability one). Spatial point patterns arise as the natural sampling information in many problems. Examples include the positions of trees in a forest, galaxies in the sky, a certain type of commerce in a city or cases of a certain disease. Seminal books on the theory of point processes and their applications include \cite{stoyan1988stochastic,Stoyanetal94,baddeley2007spatial,illian2008statistical, diggle2013statistical,cressie2015statistics}.
% More recent monographs are \cite{baddeley2007spatial,illian2008statistical,diggle2013statistical,karr2017point}.

%%%%%%%%%%%%%%%%%
In mathematical terms, if $(E)^n \equiv
E \bullet E \bullet \cdots \bullet E$ ($n$ times) is the set of
$n$ elements of $E$;

We define the exponential space as
\begin{displaymath}
E_e \equiv \bigcup_{n=0}^{\infty }(E)^n.
\end{displaymath}
%Llamaremos $B^{(n)} \equiv B \cap (\re^d)^n$,
%para todo $B$ subconjunto de $\re^d_e$.
%Sea ${\cal B}^{(n)}$ la \sigalg de Borel en el espacio producto $(\re^d)^n$,
%es decir, la menor \sigalg
%que contiene los rect\'angulos $B_1 \bullet \cdots \bullet B_n$,
%con $B_i$, conjuntos de Borel en $\re^d$. Denotamos por ${\cal B}_e$
%la menor \sigalg generada por los conjuntos $B^{(n)} \in {\cal B}^{(n)}$,
%$n=0,1,...$ Al par $(\re^d_e, {\cal B}_e)$ se le llama {\it espacio
%exponencial}, \cite{CarterPrenter72}.
A point process, $\Phi$, is defined as a measurable application of a probability space ($ \Omega,{\cal A} ,{\cal P}'$) on the measurable space
$(E_e, {\cal B}_e)$. For details on the measurable exponential space see \cite{carter1972exponential}.
%%%%%%%%%%%%%%%%%%%

\vspace{0.2cm}

Accurate and well-founded methodologies for the analysis of point processes are widely used in the literature, but most of them focus on the case where only one observation of the process is available.

There are mainly two different approaches to represent and/or describe point processes: event densities and distributions and random counting measures. In this paper we focus on the second one.  Our goal is to take advantage of the relationship between random measures and RKHS-valued random variables, to work with point processes characterized by random functions in a RKHS. Thus, we explore how we can deal with more complex statistical methodologies in this space. In particular, our proposals will make it very natural and straightforward to analyze replicated point patterns, i.e., data sets consisting of several point patterns that can be considered independent replicates of the same experiment.

\vspace{0.2cm}

RKHS spaces are well known in the statistical literature, and are frequently used in the context of Machine Learning \citep{berlinet11,saitoh16} as a valuable tool in classification or regression problems on Euclidean or $L^2$ spaces or even on Riemannian manifolds. In particular, in Euclidean spaces the success of many classification algorithms is due to the use of kernel methods \citep{scholkopf2002}. However, there is hardly any literature on statistical methods when the data live in a RKHS \citep{lukic2001stochastic}, which is our case.

\vspace{0.2cm}

The theory of statistics with functional data is an important field of research in statistics. It deals with samples in which a function is observed for each individual.
The books by \cite{SilvermanRamsay05,FerratyVieu06,horvathetal12} and \cite{aneirosetal17} are key references, as are the excellent reviews by \cite{cuevas14} and \cite{goia2016introduction}. Although the theory of functional data analysis has incorporated many tools from classical parametric or nonparametric statistics, the infinite-dimensional nature of the sample space poses particular problems \citep{FerratyVieu06}, even though, in practice, one has only sampled observed curves into a finite set of observation points.\\
There are two different perspectives on functional data. The first view is that functional data are realizations of random variables taking values in a Hilbert space. The second view is that functional data are the sample trajectories of a stochastic process with smooth mean and covariance functions. There are subtle differences between the two perspectives from a theoretical point of view. \\
RKHSs are often present in the second perspective of functional data analysis \citep{preda07bis, kadri15} since the Lo\'eve-Parzen congruence \citep{aronszajn50} links a second-order stochastic process with the RKHS generated by its covariance function \citep{eubanketal08, kupresanin10}. The functional principal component directions turn out to be an orthonormal basis of the Hilbert-Schmidt covariance operator associated with the covariance kernel \citep{horvathetal12, cuevas14}. Although, in our case the functional data be by definition a sample of a variable in a Hilbert space (first point of view), our approximation will be similar to the latter.

\vspace{0.2cm}

In \cite{baddeley2015analysing}, the practical analysis of replicated point patterns is explained using the R package \textit{Spatstat}. In particular, one of the dataset included in this package will be used in this paper: the Pyramidal dataset. This dataset contains data from \cite{diggle1991analysis} and they are locations of pyramidal neurons in human
brain of 12 normal, 9 schizoaffective, and 10 schizophrenic human subjects. All our implementations were written with R \citep{R2021}, mainly using the \textit{Spatstat} and the \textit{MASS} packages.

\vspace{0.2cm}

The article is organized as follows: \\
First at all, Section~\ref{sec:PP-RKHS} concerns the theoretical concepts. Secondly,
in Section~\ref{sec:DA} the theoretical concepts are applied to the statistical analysis of replicated point patterns.
%An experimental study with synthetic figures is conducted in Section~\ref{sec:exp_study2D}.
After taht, two particular applications are detailed in Section~\ref{sec:appl1}.
Finally, conclusions are discussed in Section~\ref{sec:conclusions}.

% =============================================================================
\section{From point processes to random elements in a reproducing kernel Hilbert space} \label{sec:PP-RKHS}
% =============================================================================

As indicated in the introduction, the objective of this paper is to show a methodology that allows the study of point processes by characterizing them by means of random functions in a RKHS.  In this section we introduce the theoretical concepts necessary for this purpose. First, the definition and properties of reproducible kernel Hilbert spaces are briefly introduced. Second, we see how to embed measures in a RKHS. Thereafter, we express a point process as a random measure and discuss some theoretical results on random variables in Hilbert spaces, in general, and RKHS, in particular.

\subsection{Reproducible kernel Hilbert spaces}

A Reproducible Kernel Hilbert Space (RKHS) is a  Hilbert space of functions $f:E \rightarrow  \re $ with some practical and interesting properties.
The theory of Reproducible Kernel Hilbert Spaces was developed by \cite{aronszajn50}.

\begin{defn} Let $H$ be a Hilbert space of real-valued functions defined on $E$ and $\langle \cdot , \cdot \rangle_H $ the inner product on  $H$. A function $k:E \times \mathbb{R}^n \rightarrow \mathbb{R}$,  is said to be an reproducing kernel (rk) associated with $H$ if it satisfies:
\begin{enumerate}
\item for every $x \in E$, $k(\cdot, x) \in H$.
\item $k$ satisfies the "reproducing property"; that is, $ \forall f \in H$ and $x \in E$
\begin{eqnarray}\label{funcio7}
f(x) = \langle k( \cdot , x) , f \rangle_H  \nonumber
\end{eqnarray}
\end{enumerate}
\end{defn}
\begin{defn}
A Hilbert space of real-valued functions is a Reproducible Kernel Hilbert Space if it has a reproducing kernel (rk) associated.
\end{defn}

A RKHS can be obtained from a kernel and each kernel determines (Moore-Aronszajn theorem) a unique RKHS, denoted by $H_k$. The construction of $H_k$ is given as follows.

We consider the set $H_0$ of linear combinations:
$$H_0=\{ \sum_{i=1}^N b_i k(\cdot, x_i), N\geq 1, b_i \in \re, \ x_i  \in E \},$$
the RKHS $H$ associated with the kernel $k$ is the closure of $H_{0}$.

Given $\phi_1=\sum_{i=1}^{N_1} a_i k(\cdot, x_i)$ and $\phi_2=\sum_{i=1}^{N_2} b_i k(\cdot, y_i)$, $\langle \phi_1,\phi_2 \rangle_k=\sum_i\sum_ja_ib_jk(x_i,y_j)$

\subsection{Embedding measures in a RKHS}

The study of random measures requires sophisticated mathematical tools.  For this reason, and following \cite{berlinet11}, we first show how reproducing kernels can be used to represent measures in functional spaces starting with Dirac measures. We will then use this embedding to define inner products.

Given a compact subset $E$ of $\re^2$, and $\mathfrak{B}$ the $\sigma-$algebra of Borel of subsets of $E$, the Dirac measure $\delta_x$ is defined for $x$ in $E$ by:
\[\delta_x(A)=\left\{\begin{array}{ccl}
1&\mbox{if}&x \in A\\
0&\mbox{if}&x \notin A \end{array}\right.\]
where $A$ is a Borel set in $E$.

The mapping:
$$
\delta_x \mapsto k( \cdot,x)
$$
embeds the set of Dirac measures on $E$  in the RKHS with kernel $k$.

If the function $k(y,\cdot)$ is measurable, the value $k(y,x)$ of the function $k(\cdot,x)$ at the point $y$ can be written as the integral
$\int k(y,t)d\delta_x(t)$, and the mapping can be rewritten as:

$$
\delta_x \mapsto \mathcal{I}_{\delta_x}= \int k(y,t)d\delta_x(t)
$$

In addition, for any measurable function $f$ in $H_k$
$\langle f, k( \cdot,x) \rangle_H =f(x)=\int f(t)d\delta_x(t)$

More generally, if $x_1,...,x_N$ are $N$ distinct points in $E$ and $b_1,...,b_N$, $N$ non null real numbers, a linear combination
\begin{displaymath}
\mu=\sum_{i=1}^{N}b_i\delta_{x_i}
\end{displaymath}
of Dirac measures is called finite support signed measure. We can extend the previous mapping with
\begin{equation} \label{mapping1}
\mu \mapsto \sum_{i=1}^{N}b_{i}k(\cdot,x_i)=\int k(y,t)d\mu(t).
\end{equation}

This mapping embeds in $H_k$ the set of measures on $E$ with finite support, $\mathcal{M}_0$, and the set $H_0$ can be seen as the set of its representers in $H_k$. Again, we have the property that for any measurable function $f$ in $H_k$
$ \langle f,  \sum_{i=1}^{N}b_{ik}(\cdot,x_i) \rangle_H= \sum_{i=1}^{N}f(x_i)=\int fd\mu.$

Following the generalisation, we can also embed the set $\mathcal{M}$ of signed measures on $E$ in $H_k$ with the mapping:
\begin{eqnarray*}
\nonumber % Remove numbering (before each equation)
  \mathcal{M} &\rightarrow & H_k \\
 \mu &\mapsto & \mathcal{I}_{\mu}=\int k(\cdot,t)d\mu(t)
\end{eqnarray*}
%and $\langle \mu, \nu \rangle = \int k d\mu \otimes d\nu,$
see \cite{berlinet11} to more details.

If we assume that $k$ is such that the functions $I_{\mu}$ and $I_{\nu}$ are different if $\mu$ and $\nu$ are not equal, Theorem 99 of \cite{berlinet11} guaranties that the mapping:
\begin{eqnarray*}
\nonumber % Remove numbering (before each equation)
  \mathcal{M} \times \mathcal{M} & \rightarrow & \re \\
 (\mu, \nu) & \mapsto & \langle \mathcal{I}_{\mu}, \mathcal{I}_{\nu}\rangle_H
\end{eqnarray*}
defines an inner product on $\mathcal{M}$ for which $\mathcal{M}_0$ is dense in $\mathcal{M}$ and its converse.

\cite{guilbart1979produits} was the pioneer in studying the relationships between reproducing kernels and inner products on the space $\mathcal{M}$. He exploited the embedding and characterized the inner products inducing the weak topology on sets of measures.

\subsection{Point processes and random measures}

If $\Phi$ is a point process, the random counting measure associated to $\Phi$ is defined as:
\begin{equation} \label{pp1}
\mu_{\Phi}(B)=\# \{ x: \ x \in \Phi \cap B\},
\end{equation}
for $B$ a Borel set in $E$.

The random counting measure of a point process characterizes its probability distribution and provides a framework for developing the theory of point processes as part of a general theory of random measures \cite{daley2008basic}.

If $\mathcal{M}$ is a set of measures in $E$ equipped with some $\sigma$-algebra, a random measure can be regarded as a random variable with values in $\mathcal{M}$. For a detailed study of the theory of random measures, see
\cite{daley1998introduction}.
Although this concept seems easy, the very definition of random measures raises delicate problems. For instance, the definition of the $\sigma$-algebra possibly derived from some topology on $\mathcal{M}$ is not a simple matter and the resulting theory involves delicate mathematical questions.

For this reason, once we have seen how we can embed the set of measures in a RKHS, we use this embedding to define and study random measures as random elements in a RKHS i.e. RKHS-valued random variables.
We will assume that the random variable takes its values in $\mathcal{M}$ (or $\mathcal{M}_0$)  with probability 1. This is the construction proposed by \citep{suquet1986espaces}.

With this construction, we can subsequently exploit the known results concerning the probability laws in a separable Hilbert space. In addition, RKHS are vectorial metric spaces and they can be considered as the natural extension of the usual Euclidean spaces. Most of the theoretical and methodological statistical results defined in Euclidean spaces are directly inhered in RKHS spaces.  Furthermore,  the completeness of Hilbert spaces gives a framework in which to work with infinite-dimensional vectors as the limit of finite-dimensional vectors.

Probability theory in Banach and Hilbert spaces is an important branch of modern probability. A complete treatment of this topic can be found in \cite{ledoux1991probability} and \cite{Hsing15}. A large number of concerning large sample results can be applied on a separable Hilbert space.
Among all these theoretical results, we recall here only a central limit theorem which guarantees convergence to a Gaussian process analogous to Euclidean spaces. See \cite{Hsing15} for a complete exposition.
\begin{thm} \label{tcl}
Let $\chi_1,...,\chi_n$ be independent and identically distributed random elements in a Hilbert space $H$ with mean 0 and $ \mathbb{E}\|\chi_i\|^2 <\infty$. Then
$$\xi_n:=\frac{\sum_{i=1}^n \chi_i }{\sqrt(n)} \rightarrow^d \xi,$$
where $\xi$ is Gaussian random element of $H$ with covariance operator equal to $\mathbb{E}(\chi_i\otimes\chi_i)$.
Being $\otimes$ the tensor product operator in $H$, $\| \cdot \|$ the norm defined by the interior product in $H$ and $\rightarrow^d$ denotes convergence in distribution.
\end{thm}

Furthermore, if the Hilbert space is also a RKHS, we have more important and strong properties which guarantees the application of standard statistical techniques to random variables in a RKHS.
\cite{guilbart1979produits} proved a Glivenko-Cantelli theorem that he applied to estimation and hypothesis testing.
\cite{berlinet1980variables,berlinet1980espaces} studied weak convergence in the set of probabilities on a RKHS, measurability and integrability of RKHS-valued variables.  Another important result, although it will not be used in this paper, is the theorem 7.5.1 of \citep{Hsing15}, which tells us that a random variable in a RKHS is also a stochastic process and the reciprocal.
%\begin{thm}
%A random element $X$ of $H_k$ is a stochastic process. Conversely,
%%a stochastic process $X$ taking values in $H_k$ is a random element of
%$H_k$.
%\end{thm}
%This theorem tells us that we can also use the  stochastic processes point of view to analyse random variables in a RKHS \cite{grenander1950stochastic,cuevas14,Hsing15}.

\vspace{0.2cm}

Turning to the case of point processes, Equation \ref{pp1} can be rewritten as:
\begin{equation} \label{eq:pprkhs}
\mu_{\Phi}(B)=\sum_{x_i \in \Phi} \delta_{x_i}(B)=\sum_i^{N} \delta_{x_i}(B),
\end{equation}
for $B$ a Borel set in $E$ and where $\delta$ denotes the Dirac measure.

The counting measure defined from $\Phi$ is a random measure and all previously mentioned for random measures applies to the statistical analysis of replicated point processes.

A counting measure is a particular case of finite support measure and we can embed it in a RKHS using the mapping \ref{mapping1}:
\begin{equation} \label{mapping2}
\mu_{\Phi} \mapsto \sum_{i=1}^{N}k(\cdot,x_i).
\end{equation}
And, we will assume that its associated RKHS random variable takes its values in the set of counting measures on $E$, $\mathcal{N} \subset \mathcal{M}_0$,  with probability 1.

\vspace{0.2cm}

Moreover, it is satisfied that if two counting measures $\mu$ and $\nu$ are not equal, i.e., they have different supports $\{x_1,...,x_N\}$ and $\{y_1,...,y_M\}$,
$I_{\mu}=\sum_{i=1}^{N}k(\cdot, x_i)$ and $I_{\nu}=\sum_{i=1}^{M}k(\cdot, y_i)$ are different.

\vspace{0.2cm}

Regarding computational aspects, the mapping \ref{mapping2} can be easily obtained using, for example, the function \textit{density} of the spatstat package of R.

Although the possibilities opened up by this way of working are enormous, in this paper, we just focus on two applications for illustrative purposes.

In the first application, we have different experimental groups, we observe independent replicates of a point process within each group and we are interested in contrasting whether there are differences between groups, i.e. an ANOVA problem where the response is a point pattern.
In the second example, we also have different groups but we are interested in a rule to classify a new point pattern in one of these groups, i.e. a supervised classification problem when the explanatory variable is a point pattern.

In both cases, classical multivariate statistical methods will be used in order to take profit of all the theoretical results aforementioned.

% =============================================================================
\section{Application to the statistical analysis of replicated point processes} \label{sec:DA}
% =============================================================================

In the previous section, we have seen  how to transform point processes into random elements in a RKHS ; therefore, in this section, we assume that we have a sample $\{\varphi_i(\cdot)\}_{i=1}^n$ of a  random element in the RKHS $H_k$, each of the elements of the sample having the expression:
\begin{equation}\label{eq:span}
\varphi_i(x)=\sum_{l=1}^{N_i} k( x,x_{il}),  x \in E
\end{equation}

Because our data are a particular kind of functions, it would make sense to use the insights of functional data analysis (FDA) to carry out any statistical analysis. But, our data are very different to the typical data in FDA issues.
Firstly, our data are not expressed in the standard form, where the $i$-th functional datum is given just by a set of discrete measured values $((y_{i1},t_1),\cdots,(y_{im_i},t_{m_i}))$ with little knowledge of the analytical form of the function. Each datum  of our sample is a function $\varphi_i(x)$ whose analytical expression is defined by  (Eq.~\ref{eq:span}).
Secondly, our function space is a  RKHS, i.e. our raw functional data ``live'' in a RKHS and, as it was said before,  its vector structure and inner product can be exploited in the data analysis.
For these reasons, our proposal is to express each function of our dataset with respect to the orthonormal base given by the eigenfunction decomposition of the kernel that defines the RKHS. The elements given by the eigenfunction decomposition of the kernel are orthonormal and are ordered according to an optimality approximation criterion. These properties allow us to reduce the dimension and thus we can apply classical statistical procedures as in the multivariate Euclidean case.
Similar ideas were previously used by authors in a very different context: supervising classification of geometrical object. The following results are similar to results of Section 3 in \cite{Barahonaetal16}.

\vspace{0.2cm}

To obtain the projection on this base, we need first to change the expression of our functions so that the points at which the kernel is evaluated are the same in all the point patterns of the sample. For that we can use the following theorem.

\begin{thm}[Representer Theorem] \label{thm:representer}
Given $ \varphi_i(\cdot)$, a grid $\{a_l\}_{l=1}^{N}$ in $E$, denoting $\varphi_i(a_l)=b_{il}$ and given a regularization parameter $\gamma > 0$ then $\exists! \chi_i \colon \re^2 \rightarrow \re$; $\chi_i(y)=\sum_{l=1}^N \beta_{il} k(y, a_l)$ such as:
\begin{equation}\label{eq:representer theorem}
 \chi_i= \argmin_{g \in H_k} \frac{1}{N} \sum_{i=1}^N (g(a_l)-b_{il})^2 + \gamma \norm{g}_{H_k}^2 \text{,}
\end{equation}
where $\beta_{il} \in \re$ (for $l=1, \dots, N$) are the solutions of:
\begin{equation*}%\label{eq:funcio9}
(\gamma \, N \, {\mathbb I}_{N\times N} +K|_a)\beta_i= b_i,
\end{equation*}
with $K|_a$ the matrix defined as $(K|_a)(i,j)=k(a_i, a_j)$, $i,j=1,\dots, N$, and $\beta_i$, $b_i$ are the $N \times 1$ vectors
\begin{equation*}
\beta_i=
\begin{pmatrix}
\beta_{i1} \\
\beta_{i2} \\
\vdots  \\
\beta_{iN}
\end{pmatrix}
;\quad b_i=
\begin{pmatrix}
b_{i1} \\
b_{i2} \\
\vdots \\
b_{iN}
\end{pmatrix}
\end{equation*}
\end{thm}
As a result of applying the theorem,  from now on we will work with the sample:
\begin{equation} \label{Eq:chi}
\{ \chi_i(\cdot)=\sum_{l=1}^N \beta_{il} k(\cdot, a_l )\}_{i=1}^n
\end{equation}

\vspace{0.2cm}

It is well known \citep{Hsing15} that if $E$ is compact and $k$ continuous, measurable and bounded, the integral operator associated to the kernel function $k$ and defined by:
\begin{equation}
(\mathcal{K}f)(\cdot)=\int_{E}k(\cdot, x )f(x)dx,  \ \   f \in \mathcal{L}_2(E),
\end{equation}
where $\mathcal{L}_2(E)$ is the space of square integrable functions on $E$, is a compact, continuous, self-adjoint, and positive operator.

As a result,  it can be expressed as
$$ \mathcal{K}=\sum_{q\geq1}\lambda_q e_q \otimes e_q, $$ where $\{\lambda_q, e_q \}_q$ is the countable sequence of its eigenvalues and (orthonormal) eigenfunctions (spectral decomposition).

In addition:
$$ k(x,y)=\sum_{q\geq1} \lambda_q e_q (x) e_q (y) $$.

And $\{\sqrt{\lambda_q} e_q \}_q$ is a complete orthonormal basis for $H_k$.

The following results show us how we can project our sample in this base of the RKHS.

\begin{thm}\label{thm:tours}
Let $\chi_i$ be an element of the RKHS $H_k$, it can be expressed as
\begin{equation}\label{eq:expresion_base}
\chi_i(\cdot)=\displaystyle \sum_{q=1}^{\infty}\mu_{qi} \left(\sqrt{\lambda_q}e_q (\cdot)\right),
\end{equation}
where $\{ \sqrt{\lambda_q} e_q \}_{q=1}^{\infty}$.

In addition, if $\chi_i(\cdot)=\displaystyle \sum_{q=1}^{\infty}\mu_{qi} \left(\sqrt{\lambda_q}e_q (\cdot)\right)$ and $\chi_j(\cdot)=\displaystyle \sum_{q=1}^{\infty}\mu_{qj} \left(\sqrt{\lambda_q}e_q (\cdot)\right)$, the inner product is:
$$\langle \varphi_i, \varphi_j\rangle_k=\sum_q \mu_{qi}\mu_{qj}.$$

Furthermore \citep{smale09,GonzalezMunoz10}, the first $d= \rank(K|_a)$ coefficients $\mu_{qi}$ can be approximated by
\begin{equation}\label{eq:funcio14}
\widehat{\mu_{qi}}=\sqrt{\ell_q}(v_q \cdot \beta_i)
\end{equation}
where $v_q \in \re^N$ are the eigenvectors of $K|_a$, $\ell_q$ are the eigenvalues of $K|_a$.
\end{thm}

The following result assures us that if we express our infinite-dimensional data in this basis and truncate it to obtain a finite-dimensional vector sample, we will have the highest possible accuracy.

\begin{prop}
Theorems~4.4.7 and 4.6.8 in \cite{Hsing15} tell us that for a fixed integer $r>0$ with $\lambda_r >0$:
\begin{equation*}
\min_{f_1,\dots,f_r\in H_k} \int\int_{E \times E}\left( k(y,x)- \sum_{q=1}^{r} f_{q}(y)f_{q}(x)\right)^2 dydx,
%= \sum_{q=r+1}^\infty \lambda_q^2,
\end{equation*}
%where
the minimum is achieved by $f_q=e_q$.
\end{prop}

This result ensures that the truncated eigenvalue-eigenvector decomposition provides the best approximation to $k$ and, as a result, to our data (Eq. \ref{Eq:chi}).

If we truncate the summation in Equation \ref{eq:expresion_base} to a low number of terms $r$, $r\leq d= \rank(K|_a)$, each point pattern $\varphi_i$ for $i=1, \dots, n$, is given by the coefficients $\mu_{qi}$ for  $q=1, \dots, r$ (estimated by $\widehat{\mu_{qi}}$), on the orthonormal basis $\{ \sqrt{\lambda_q} \psi_q \}_{q=1}^{\infty}$. As a result, it can be represented as the $ r$-dimensional vector
\begin{equation}%\label{eq:fapprox}
\mu_i= (\widehat{\mu_{1i}}, \widehat{\mu_{2i}},  \dots , \widehat{\mu_{ri}}).
\end{equation}

This expression optimally reduces the infinite-dimensional problem to a finite-dimensional problem. It is now possible to apply well-known classical multivariate methods.
In particular, in the following section two well-known classical multivariate methods will be used for illustrative purposes: MANOVA and Discriminant Analysis.

% =============================================================================
\section{Application to  locations of pyramidal neurons in human
brain }\label{sec:appl1}
% =============================================================================

\begin{figure}[htbp]
\begin{center}
\begin{tabular}{cc}
\includegraphics[width=8cm]{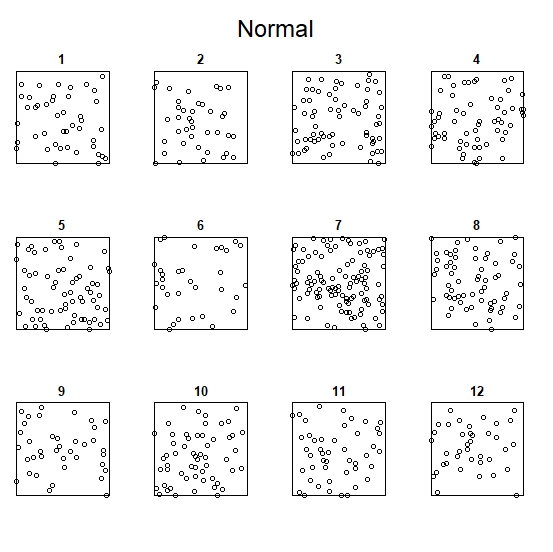} &
\includegraphics[width=8cm]{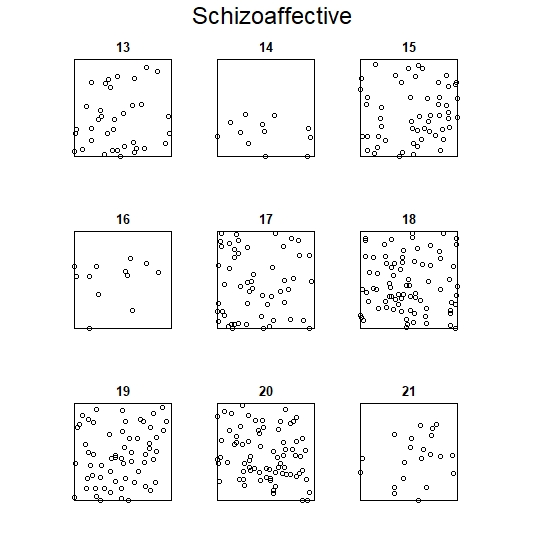} \\
\includegraphics[width=8cm]{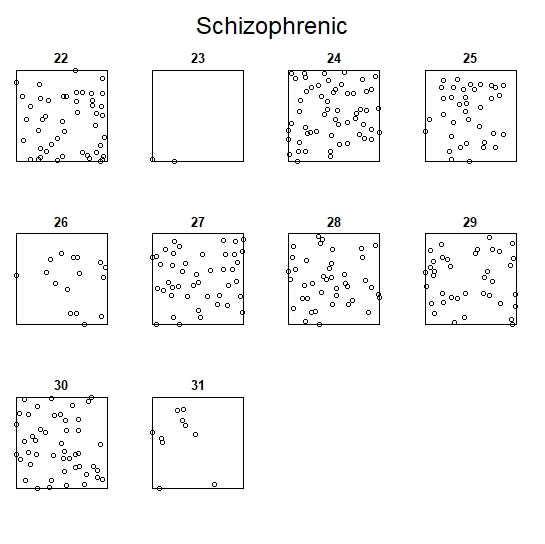} &\\
\end{tabular}
\caption{Positions of the pyramidal neurons for the three groups of subjects. \label{fig:pyramidal}}
\end{center}
\end{figure}

As mentioned before, the \textit{Pyramidal dataset} is included in the R package \textit{spastat} and contains data from \cite{diggle1991analysis}. The data are the locations of pyramidal neurons in human brain. One point pattern was observed in each of 31 human subjects.
There were 12 normal control, 9 schizoaffective, and 10 schizophrenic cases. See \cite{diggle1991analysis} for a more detailed explanation of them.
Figure \ref{fig:pyramidal} gives plots of the point patterns in our sample, one plot for each subject.

In this example $E=[0,1] \times [0,1]$ and a gaussian kernel with $\sigma=0.05$ were used to embed the point patterns in a RKHS using the map \ref{mapping2}. A grid $\{a_l\}_{l=1}^{N}$ with $a_l$ equally spaced in $E$ with a vertical and horizontal step of $h=0.02$ is used to apply the representer theorem \ref{thm:representer}. The parameter $\gamma=0.000127$ in Equation \ref{eq:representer theorem} is fixed to the minimum value wich makes matrix $(\gamma \, N \, {\mathbb I}_{N\times N} +K|_a)$ invertible. The parameters $h$ and $\sigma$ are related with the accuracy of our working data. When $h$ and $\sigma$ decrease, we have more accuracy, but more computational cost and complications in the calculations. The chosen values represent a balance between both.

After applying the representer theorem we have a sample $ \{\chi_i(\cdot), y_i \}_{i=1}^{31}$ with $\chi_i(\cdot)=\sum_{l=1}^N \beta_{i,l} k(\cdot, a_l )$, and $y_i$ a categorical variable indicating if the subject is normal, schizoaffective or schizophrenic.

\begin{figure}[htbp]
\begin{center}
\begin{tabular}{ccc}
\includegraphics[width=4.4cm]{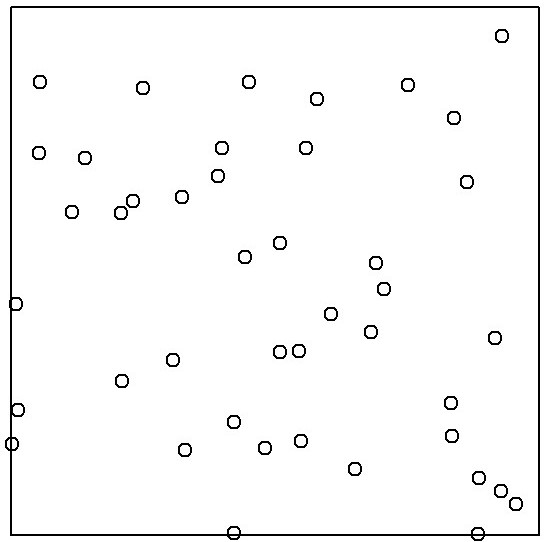} &\includegraphics[width=4.4cm]{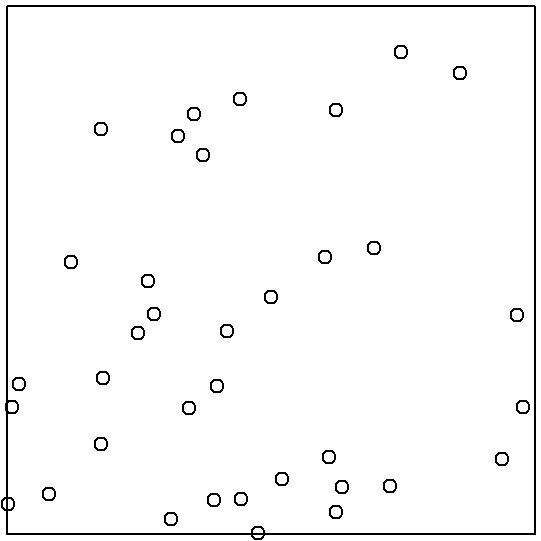} &\includegraphics[width=4.4cm]{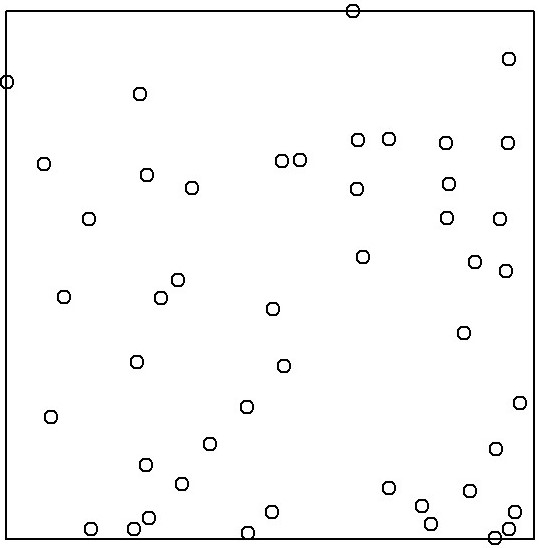}
\\
\includegraphics[width=4.6cm]{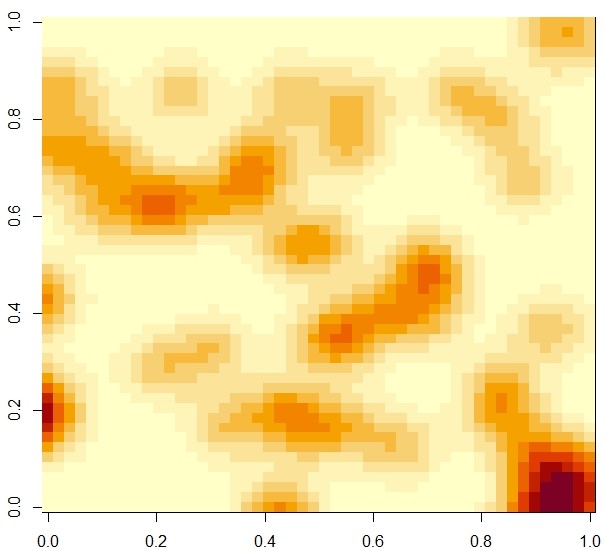} &\includegraphics[width=4.5cm]{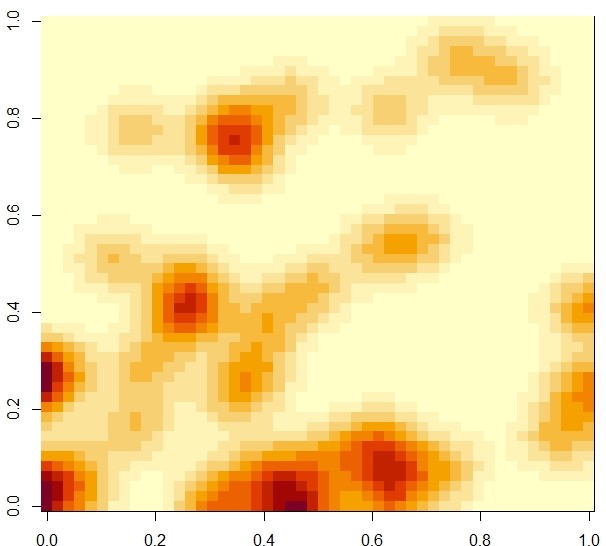} &\includegraphics[width=4.5cm]{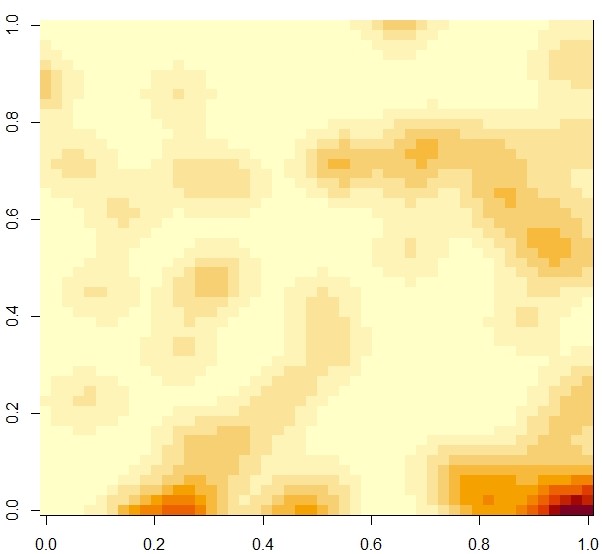}
\\
\includegraphics[width=4.5cm]{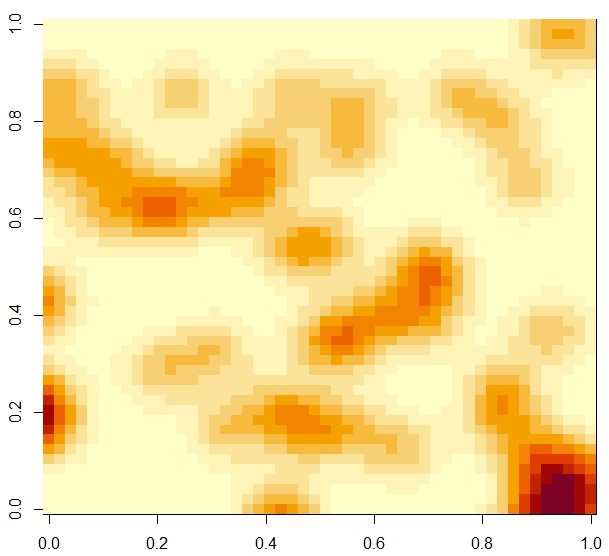} &\includegraphics[width=5cm]{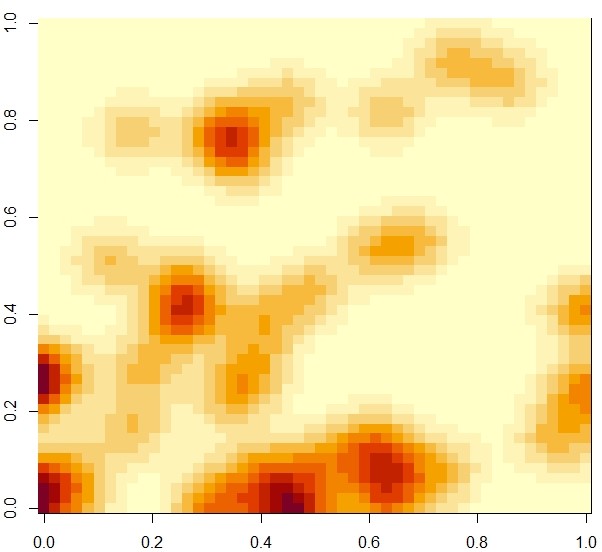} &\includegraphics[width=4.5cm]{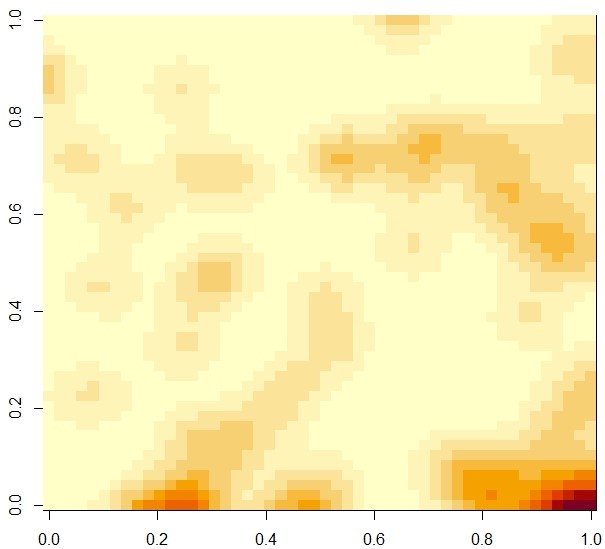}
\\
\end{tabular}
\caption{Point pattern and RKHS functions of the first subject of each class before and after applying the representer theorem. \label{fig:aplication11}}
\end{center}
\end{figure}

In Figure \ref{fig:aplication11} we can see the plot of one point pattern of each group and its corresponding elements $\varphi_i$ and $\chi_i$ in the RKHS.

Finally, Equation~\ref{eq:expresion_base} is applied to obtain the vector of coefficients $\mu_i$ that represent each point pattern in relation to the orthonormal base of $H_k$. Because the size of our sample is very small (12, 9 and 10 cases in each group), we truncate it at $r=6$.

\begin{figure}[htbp]
\begin{center}
\includegraphics[width=15cm]{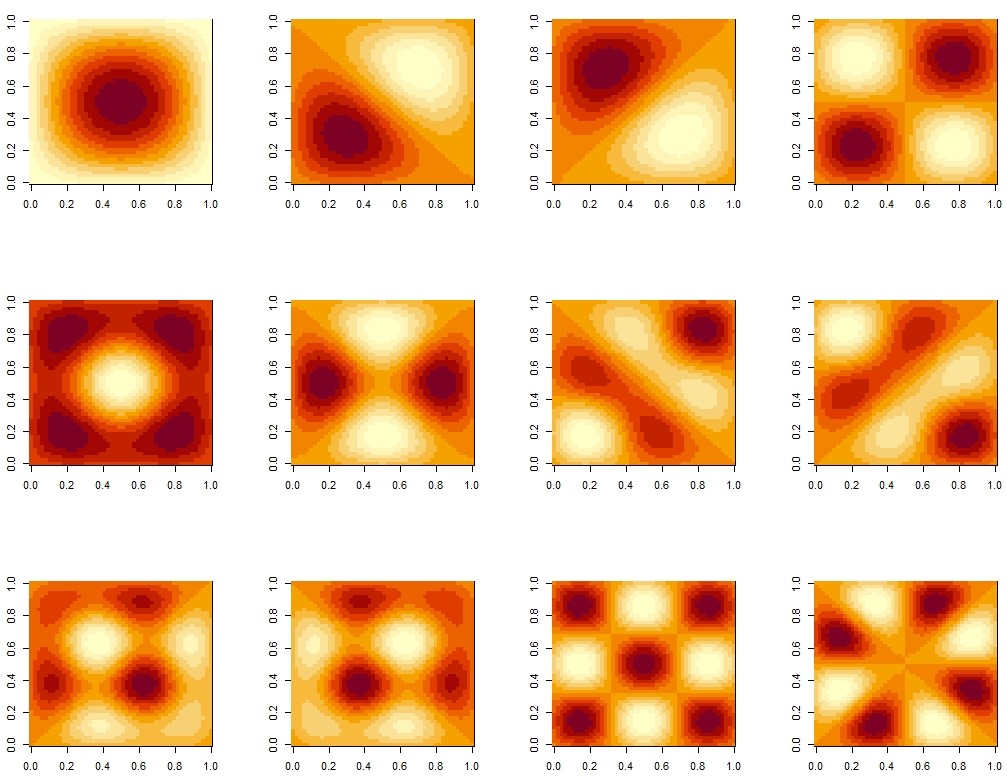}
\caption{Firsts eigen functions of the Gaussian Kernel in a $[0,1] \times [0,1]$ window. \label{fig:eigenfunctions}}
\end{center}
\end{figure}

It is worth analysing the meaning of the firsts functions of the base.
In Figure~\ref{fig:eigenfunctions}, we can see the six firsts eigen functions.
The coefficient of the first function measure high density concentrated in the center of the window, the second and third high density just within one half of the window and so on.

%=================================================================================================
\subsection{ANOVA}
%=================================================================================================

Two principal approaches can be found in the literature of spatial point patterns to identify significant differences between several experimental groups  \citep{diggle2000comparison}.
The first one \citep{diggle1991analysis, ramon2016new, gonzalez2018statistical} is based on functional descriptors of the pattern and nonparametric inference.

The second one is based on assuming a parametric model for the pattern (usually pairwise interaction point process or Gibbs processes), the parameter of the models are estimated using maximum likelihood or pseudo-likelihood methods for each group and differences between groups are tested by comparing fits with and without the assumption of common parameter \citep{illian2010gibbs,illian2012toolbox}.

Following the aforementioned procedure,  we will work with the RKHS elements resulting of the embedding.

\begin{figure}[htbp]
\begin{center}
\begin{tabular}{ccc}
\includegraphics[width=4.5cm]{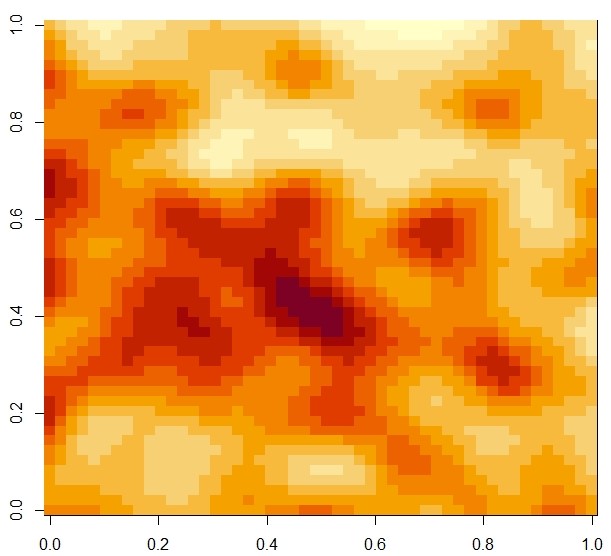} &\includegraphics[width=4.5cm]{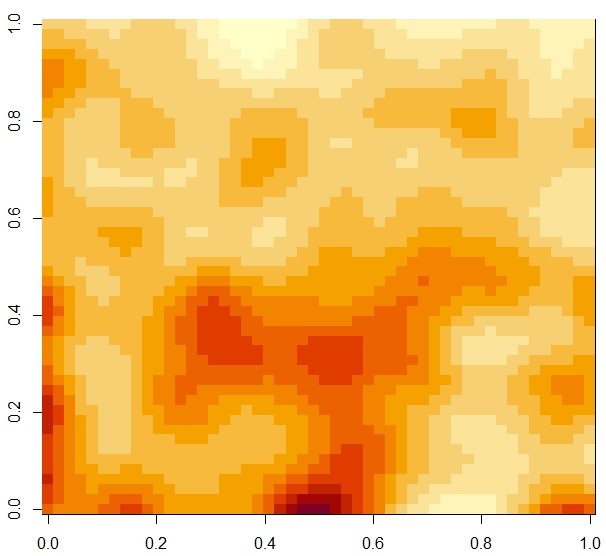} &\includegraphics[width=4.5cm]{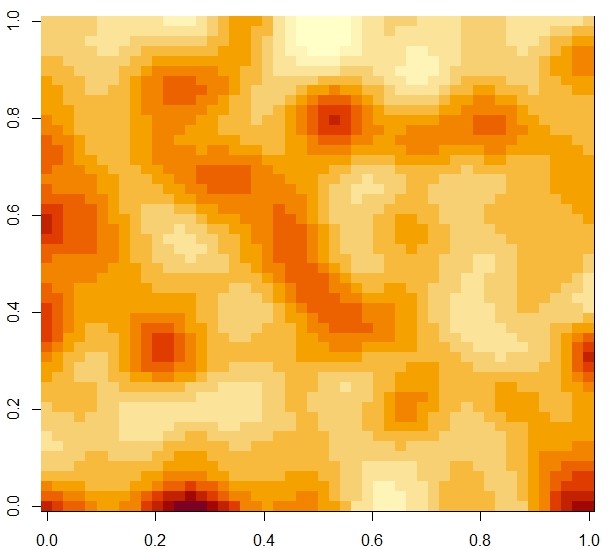}\\
\end{tabular}
\caption{The means of the RKHS functions of each class: normal, schizoaffective and schizophrenic.}
\label{fig:aplication12}
\end{center}
\end{figure}

Since a RKHS is a vector space, the mean sample element is simply obtained as:
$\bar{\chi}(\cdot)=\sum_{l=1}^N \sum_{i}^n \frac{\beta_{i,l}}{n}  k(\cdot, a_l )$. In Figure~\ref{fig:aplication12} we can see the mean sample element of each group.

In this example, we are only focused on the answer to the following question: do the observed patterns differ significantly in mean from group to group?  For this purpose, a Multivariate ANOVA test can be applied to the $r$-variate sample $ \{\mu_i, y_i \}_{i=1}^{31}$.
Before the MANOVA test, a Box's M test must be applied to test the assumption of homogeneity of variance-covariance matrices. The results of both tests can be found in Table~\ref{tab:tablaanova}. At a significance level of 0.05, the hypothesis of homogeneity of variance-covariance matrices would be accepted and the hypothesis of equality of means would be rejected.

It is important to note that, in this case, the sample size is small and therefore, the Central Limit Theorem is not applicable to assume normality. Shapiro univariate tests applied to the residuals gave non significative results. Although these results do not guarantee multivariate normality, is not a cause for concern for two reasons. The first reason is the robustness to non-normality of the MANOVA test, and the second is that this example is only illustrative.

Once the multivariate hypothesis of equality of means has been rejected, we perform an univariate ANOVA test to analyse with more detail the differences. The respective p-values for the first six coefficients of our base were:  0.069,  0.044, 0.066, 0.42, 0.58 and 0.115, and we can conclude that the differences between classes are mainly in the four coefficients corresponding to the four first functions of  Figure~\ref{fig:eigenfunctions}.

It is important to emphasize again that, our objective in this section is to show the potential applications of the proposed methodology, a deeper study conducted in conjunction with experts in neuroanatomy would be necessary to reach clinical conclusions.

\begin{table}[htbp]
\begin{center}
\begin{tabular}{|c|c|c|c|}
\hline
 Test & Statistic &    df &  p-value \\
 \hline
Box's M-test &  36.009        &   42  & 0.7304\\
Manova  &    2.2358      &  12-48   &0.02451\\

\hline
\end{tabular}
\caption{Box's M and MANOVA results.}
\label{tab:tablaanova}
\end{center}
\end{table}

% ======================================================================================================
\subsection{Discriminant Analysis }\label{sec:appl2}
% =========================================================================================================

Once significant differences between groups of subjects have been found, it could be very useful to find a decision rule to classify a new individual as normal, schizoaffective or schizophrenic on the basis of its spatial pattern.

To our knowledge, the literature about supervised classification methods applied to replicated point patterns is scarce and relies mainly on dissimilarity-based methods \citep{mateu2015measures, pawlasova2022supervised}. Similar methods have been used for supervised classification of germ-grain models in \cite{gallego2016inhomogeneous}.

As it is well known, the literature on supervised classification methods is extensive and covers a large number of methods ranging, from those based on multivariate statistics to the most modern deep learning techniques \citep{hastie2020elements}. But, as mentioned before, in this paper, we focus only on classical multivariate statistical methods, since the theorem~\ref{tcl} would allow us to take advantage of probability parametric models as well as convergence theorems. For this reason, linear and quadratic discriminant functions \citep{venables2013modern} will be used to illustrate the new methodology proposed in this work. As it is known, both are parametric Bayesian methods that assume a multivariate Gaussian model for the explanatory variables, with and without equality of variance-covariance matrices respectively. The \textit{a priori} probabilities of each class will be estimated from the sample.

Since our dataset is very limited, we will first present some illustrations using simulated point patterns. Two different experiments were performed.

\begin{figure}[htbp]
\begin{center}
\begin{tabular}{cccc}
\includegraphics[width=3cm]{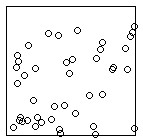} &\includegraphics[width=3cm]{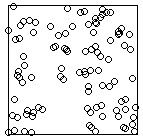} & \includegraphics[width=3.1cm]{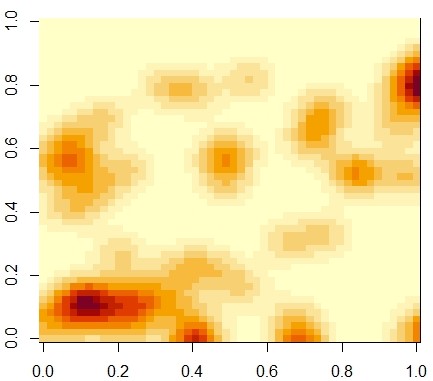} & \includegraphics[width=3.1cm]{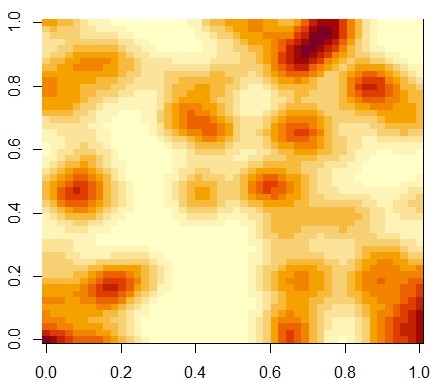}\\
\includegraphics[width=3cm]{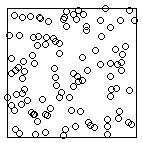} &\includegraphics[width=3cm]{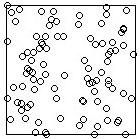} & \includegraphics[width=3.1cm]{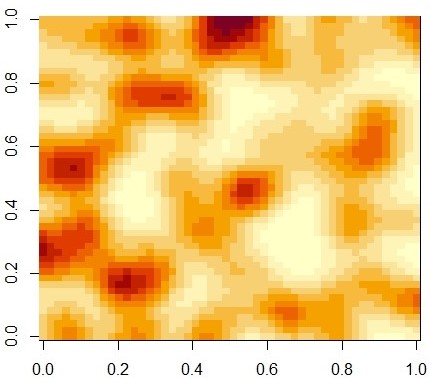}&\includegraphics[width=3.1cm]{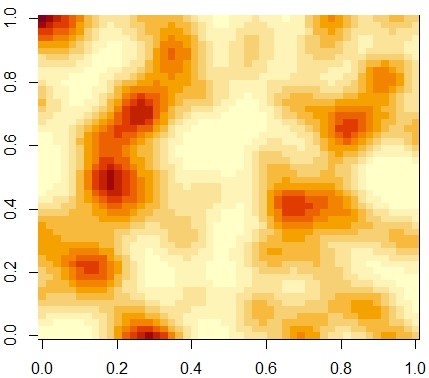}\\\
\includegraphics[width=3cm]{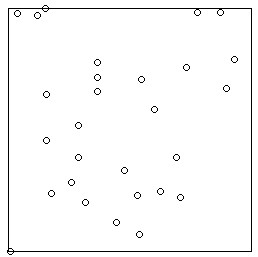} &\includegraphics[width=3cm]{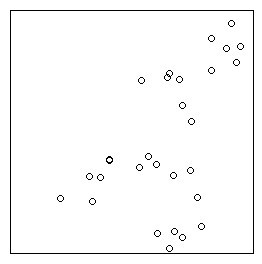} & \includegraphics[width=3.1cm]{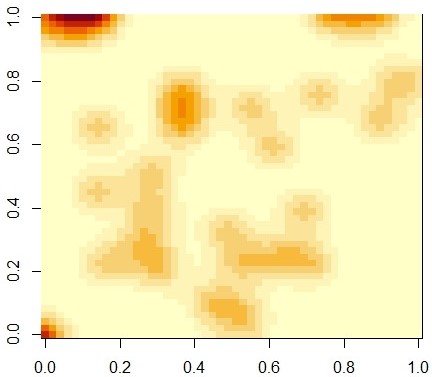}&\includegraphics[width=3.1cm]{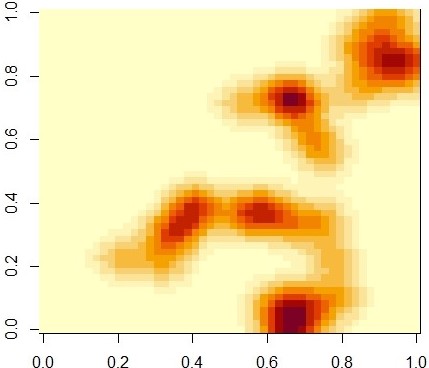}\\
\end{tabular}
\caption{One simulated point patterns and its corresponding RKHS element of each class. First row: homogeneous Poisson point processes with $\lambda_1=50$  and $\lambda_2=100$, second row: homogeneous Poisson point processes with $\lambda_1=90$  and $\lambda_2=100$. Third row: homogeneous Poisson point process with $\lambda_1=36$ and Poisson cluster point process with $\lambda_2=36$. }
\label{fig:simulated}
\end{center}
\end{figure}

In the first experiment two samples of size 20 of an homogeneous Poisson point process (HPPP) with different intensities, $\lambda_1$ and $\lambda_2$, were simulated in a rectangular window of size one. This experiment was repeated two times with $\lambda_1=50$ and $\lambda_1=90$, respectively and $\lambda_2=100$ in both cases. In Figure~\ref{fig:simulated}, we  can see one point pattern of each class and its corresponding element in the RKHS. Since the difference between the two classes of point processes is in the mean of its counting measures, linear discriminant functions have been used in both cases (\textit{lda} function of R package \textit{MASS}).  The Table~\ref{resultadosEjemplo2} shows the training errors and the cross-validation errors, excellent results have been obtained, even in the second case in which the difference between both models is very slight.

\vspace{0.2cm}

In the second experiment, two samples of size 30 of two different point processes were simulated again in a $[0,1] \times [0,1]$ window. The first sample corresponds to a homogeneous Poisson point process with intensity $\lambda_1=36$ and the second to a Poisson cluster point process (PCPP) with the intensity of the Poisson process of centres $\kappa=6$, and each cluster consisting of 6 points in a disc of radius 0.2. The resulting intensity is also $\lambda_2=36$. In Figure~\ref{fig:simulated}, we can see again one point pattern of each class and its corresponding element in the RKHS. In this case, both point processes have the same intensity and the difference between both classes is given by the spatial variability, for this reason the linear discriminant function does not work well and a quadratic discriminant function must be used (\textit{qda} function of R package MASS).

The same accuracy parameters of the previous section have been used, i.e. $\sigma=0.02$ and $h=0.05$.  Regarding to the number of variables, i.e. the value of $r$, different values ranging from $r$=6 to $r$=10 have been tested, but no major differences have been found. In general, the best results were obtained for $r$=7, and these are reported in Table~\ref{resultadosEjemplo2}. The results are again excellent.  In addition to obtaining very small cross-validation errors, the \textit{a posteriori} probabilities of well-classified cases are all practically greater than 0.95 and those of misclassified cases lower than 0.6.

\begin{table}[htbp]
\begin{center}
\begin{tabular}{|c|c|c|c|}
\hline
 Models & Intensisties  & Training error &  CV  error  \\
\hline
 HPPP-HPPP  & $\lambda_1=50$, $\lambda_2=100$        &   0           &0\\
 HPPP-HPPP & $\lambda_1=90$, $\lambda_2=100$        &    0.1          & 0.1755 \\
 HPPP-PCPP  & $\lambda_1=36$, $\lambda_2=36$      & 0.05       &   0.117    \\
\hline
\end{tabular}
\caption{Supervised classification errors in the simulated experiments.}
\label{resultadosEjemplo2}
\end{center}
\end{table}

After testing the performance of the proposed methodology on a supervised classification problem, the \textit{lda} function was applied to our real example to find a decision rule to classify a new individual as normal, schizoaffective or schizophrenic.
We used linear discriminant analysis because the Box's M test, used to test the hypothesis of homogeneity of variances in the previous section, did not yield significant differences. As expected, no good results were obtained due to the limitations of the sample. The training error was relatively small (0.29), but the cv error was not at all satisfactory (0.6).
A larger dataset would be necessary in order to be able to use in clinical practice. In our view, our methodology could be used
without modifications.

% =============================================================================
\section{Conclusions} \label{sec:conclusions}
% =============================================================================

We have introduced a new methodology for the statistical analysis of replicated spatial point patterns. This methodology is based on the fact that the probability distribution of a point process is completely determined by its associated random counting measure.
Random measures can be embedded in a RKHS and, in this way,  we transform the point process in a random element in a RKHS, where theoretical founded methods and algorithms can be applied, similar to what is done in an Euclidean space. To do so, we express our data in the base given by the kernel's eigenfunctions and truncate this expression in the required dimension. This guarantees to move to a lower dimension with the least loss of accuracy.

As an example of the potential real-life applications of the proposed methodology, we have used it to detect differences between point patterns of pyramidal neuron locations in the human brain from three groups of subjects (\cite{diggle1991analysis}. We have also used it to classify new observations using several simulated datasets.
With the results of these experiments, it can be stated that our methodology is feasible for applications.

\end{document}